\documentclass[aps,prd,superscriptaddress,nofootinbib,amsmath,amsfonts,preprintnumbers,notitlepage,10pt,english]{revtex4}
\usepackage{amsmath}
\usepackage{amssymb}
\usepackage{babel}

\makeatletter
\newcommand{\f}[2]{\frac{#1}{#2}}
\newcommand{\mk}[1]{\left( #1 \right)}
\newcommand{\kk}[1]{\left[ #1 \right]}

\newcommand{\be}{\begin{equation}}
\newcommand{\ee}{\end{equation}}
\newcommand{\bea}{\begin{eqnarray}}
\newcommand{\eea}{\end{eqnarray}}
\def\L{{\cal L}}
\def\o{{\rm odd}}
\def\e{{\rm even}}

\@ifundefined{textcolor}{}
{%
 \definecolor{BLACK}{gray}{0}
 \definecolor{WHITE}{gray}{1}
 \definecolor{RED}{rgb}{1,0,0}
 \definecolor{GREEN}{rgb}{0,1,0}
 \definecolor{BLUE}{rgb}{0,0,1}
 \definecolor{CYAN}{cmyk}{1,0,0,0}
 \definecolor{MAGENTA}{cmyk}{0,1,0,0}
 \definecolor{YELLOW}{cmyk}{0,0,1,0}
 }

\@ifundefined{textcolor}{}{%
 \definecolor{BLACK}{gray}{0}
 \definecolor{WHITE}{gray}{1}
 \definecolor{RED}{rgb}{1,0,0}
 \definecolor{GREEN}{rgb}{0,1,0}
 \definecolor{BLUE}{rgb}{0,0,1}
 \definecolor{CYAN}{cmyk}{1,0,0,0}
 \definecolor{MAGENTA}{cmyk}{0,1,0,0}
 \definecolor{YELLOW}{cmyk}{0,0,1,0}
 }

\makeatother

\begin{document}

\preprint{RESCEU-32/11}

\title{Black hole perturbation in nondynamical and dynamical Chern-Simons gravity}

\author{Hayato Motohashi}
\affiliation{Department of Physics, Graduate School of Science,
The University of Tokyo, Tokyo 113-0033, Japan}
\affiliation{Research Center for the Early Universe (RESCEU),
Graduate School of Science, The University of Tokyo, Tokyo 113-0033, Japan}

\author{Teruaki Suyama}
\affiliation{Research Center for the Early Universe (RESCEU),
Graduate School of Science, The University of Tokyo, Tokyo 113-0033, Japan}

\begin{abstract}
Chern-Simons gravitational theories are extensions of general relativity in which the 
parity is violated due to the Chern-Simons term.
We study linear perturbations on the static and spherically symmetric background spacetime
both for nondynamical and dynamical Chern-Simons theories.
We do not make an assumption that the background Chern-Simons scalar field vanishes,
which has been adopted in the literature.
By eliminating nondynamical variables using their constraint equations, 
we derive the reduced second order action from which a set of closed evolution equations containing 
only dynamical variables are immediately obtained and therefore the number of propagating 
degrees of freedom as well.
It is found that ghost is present both for the nondynamical case and 
for the dynamical case unless the background Chern-Simons scalar field vanishes.
It is also found that if the background scalar field vanishes, ghost degrees of freedom
are killed and all the modes propagate at the speed of light.
\end{abstract}

\date{\today}

\maketitle

\section{Introduction}

Black holes(BHs) and compact stars are unique and ideal places to
test the theories of gravity in the strong gravitational field regime.
For instance, oscillations of spacetime around BHs reflects underlying theory of gravity 
and carry direct informations of the structure of spacetime.
Detection of the gravitational waves coming from such regions therefore enables
us to reveal the nature of gravity.
There are other independent observational ways to probe the strong gravity regime,
such as black hole shadows, strong gravitational lensing etc.
These observations make us expect that a forthcoming decade would be an era of the test of general relativity(GR) 
in strong gravitational field regime.
Until now, many alternative theories of gravity such as $f(R)$ theories (for recent reviews of $f(R)$ theories, see e.g. Refs.~\cite{Copeland:2006wr,amendola,Sotiriou:2008rp,DeFelice:2010aj}) have been proposed and studied to understand theoretically what kinds of different phenomena are expected in such theories \cite{Yagi:2009zm,Yagi:2009zz,Starobinsky:1980te,Capozziello:2002rd,Carroll:2003wy,Bean:2006up,Carloni:2007yv,Appleby:2007vb,Starobinsky:2007hu,Motohashi:2009qn,Motohashi:2010tb,Motohashi:2010sj,Motohashi:2011wy}.

Chern-Simons(CS) gravity theories are alternative theories of gravity which violate parity
due to the so-called CS term, or the Pontryagin density, ${\tilde R}R \equiv \frac{1}{2} \epsilon_{\alpha \beta \gamma \delta} R^{\alpha \beta}_{~~~\mu \nu} R^{\gamma \delta \mu \nu}$, where $\epsilon_{\alpha \beta \gamma \delta}$ is the totally antisymmetric tensor.
The presence of the $\epsilon_{\alpha \beta \gamma \delta}$ tensor manifests the parity violation.
Because the CS term can be expressed as a divergence, simply adding $C$ into the Einstein-Hilbert action amounts to an addition of a total derivative term in the action and does not change the theory.
A gravitational theory in which the CS term is coupled to an external scalar function was introduced in Ref.~\cite{Jackiw:2003pm}.
In this so-called nondynamical CS theory,
a kinetic term for the external field is absent and a functional form 
of the external field is assumed by hand.
In the so-called dynamical CS theories, the external scalar function is promoted to a dynamical field \cite{Smith:2007jm}
by giving a kinetic term for the CS scalar field.
In this case, the functional form of the CS scalar field is no longer given by hand 
and must be obtained by solving the dynamical field equations. 
Many studies based on the dynamical CS theories have appeared \cite{Yunes:2010yf,Pani:2011xj,Cambiaso:2010un,Garfinkle:2010zx,Molina:2010fb,Yunes:2009ch,Cardoso:2009pk,Sopuerta:2009iy,Yunes:2009hc,Konno:2009kg,Yunes:2008ua,Konno:2008np,Yunes:2007ss,Grumiller:2007rv,Konno:2007ze}. 
For a recent review on the Chern-Simons gravity, see Ref.~\cite{Alexander:2009tp}.

A noticeable feature of the CS term is that it identically vanishes for the spherically symmetric metric.
Because of this, as long as we are looking at the spherically symmetric metric such as the one around
the sun or the static BH, there is no way to probe the effects of the CS term.
However, things drastically change if we consider oscillations of the BHs where the CS term generically 
no longer vanishes and enters the game.
Behavior of the BHs oscillations reflects the presence of the CS term and therefore
we can probe parity violation in gravity sector by seeing how the BHs oscillate.
This is the reason why we consider BH perturbation for the CS gravitational theories 
deserves investigation.

The BH perturbation for the nondynamical CS theories has been studied in Ref.~\cite{Yunes:2007ss}.
It was found there that the odd-type and even-type perturbations are generically coupled,
which is a distinct feature of the parity violation.
It was also found that purely odd or purely even perturbations are forbidden.
In particular, the so-called Cunningham-Price-Moncrief variable, which represents
partial excitation of the odd modes, is forced to vanish,
implying that the odd-type perturbations get suppression compared to the case of GR.
However, since the basic perturbation equations are highly involved, 
issues such as construction of the closed evolution equations that contain only dynamical variables 
and the number of degrees of freedom for the perturbations have remained unsolved.

The BH perturbation for the dynamical CS theories has been also addressed in Ref.~\cite{Yunes:2007ss}.
Later, authors of Ref.~\cite{Cardoso:2009pk,Molina:2010fb} showed that if the CS scalar field vanishes on
the Schwarzschild background spacetime, the gravitational odd and even-type perturbations decouple
and the effects of the CS term only appear in the odd sector.
Under an assumption that the background CS scalar vanishes, they derived closed set of equations
that contain only dynamical variables and thus are convenient for numerical integration
of the perturbation equations.
Studies of the BH perturbation for other theories of gravity include Ref.~\cite{DeFelice:2011ka}
for $f(R,{\cal G})$ (${\cal G}$ is the Gauss-Bonnet term) theories,
Ref.~\cite{Motohashi:2011pw} for $f(R,{\tilde R}R)$ theories and Ref.~\cite{Moon:2011fw} for 
$f(R)$ gravity with the dynamical CS term.

In this paper, we extend the previous studies \cite{Yunes:2007ss,Cardoso:2009pk,Molina:2010fb} on the BH perturbations for nondynamical
and dynamical CS theories to address issues that have not been clarified yet.
These include clarifications of a number of propagating degrees of freedom and the presence or 
absence of ghost for the nondynamical and the dynamical cases with
nonvanishing background scalar field.
To this end, we use second order action for the perturbation variables rather than resorting to 
the perturbed field equations themselves, which enables us to address the
ghost issue and to easily implement elimination of the nondynamical variables. 
This paper reports on the results of this calculation.
It is found that ghost is generically present both for the nondynamical and the dynamical 
cases with nonvanishing background scalar field,
which therefore puts strong constraint on the general CS theories.
It is also found that if the background scalar field vanishes, ghost degrees of freedom
are killed and all the modes propagate at the speed of light.

The organization of this paper is as follows.
In Sec.~\ref{non-d-CS}, we will develop linear perturbation analysis for the nondynamical CS theories.
In Sec.~\ref{d-CS}, we will develop linear perturbation analysis for the dynamical CS theories.
The last section is the conclusion.
We use metric signature $- + + +$ and define the Riemann tensor and Ricci tensor as
$R^\alpha_{~\beta \gamma \delta}=\partial_\gamma \Gamma^\alpha_{~\beta \delta}-\partial_\delta \Gamma^\alpha_{~\beta \gamma}+\Gamma^\alpha_{~\lambda \gamma} \Gamma^\lambda_{~\beta \delta}-\Gamma^\alpha_{~\lambda \delta} \Gamma^\lambda_{~\beta \gamma}$ and $R_{\mu \nu}=R^\alpha_{~\mu \alpha \nu}$.

\section{BH Perturbations in Non-dynamical Chern-Simons theory}
\label{non-d-CS}
\subsection{Action and field equations}
The action for the nondynamical Chern-Simons theory is given by \cite{Jackiw:2003pm}, 
\begin{equation}
S_{\rm CS}=\frac{M_P^2}{2} \int d^4x~\sqrt{-g}\left( R-\frac{1}{4} \psi {\tilde R}R \right). \label{act1}
\end{equation}
Here, $M_P=1/\sqrt{8\pi G_N} \simeq 4.34 \times 10^{-6} {\rm g}$ is the reduced Planck mass
and the CS term is given by 
${\tilde R}R \equiv \frac{1}{2} \epsilon_{\alpha \beta \gamma \delta} R^{\alpha \beta}_{~~~\mu \nu} R^{\gamma \delta \mu \nu}$ 
and $\psi$ is a CS scalar field that has dimension of ${\rm (length)}^2$.
$\epsilon_{\alpha \beta \gamma \delta}=\sqrt{-g} \eta^{\alpha \beta \gamma \delta}$ is 
the totally antisymmetric tensor with $\eta^{0123}=1$. 
By taking variation with respect to $g_{\mu \nu}$, we get gravitational field equations
which are given by
\begin{equation}
G_{\mu \nu}+C_{\mu \nu}=0, \label{NDCD-basic-eq}
\end{equation}
where $G_{\mu \nu}$ is the Einstein tensor and $C_{\mu \nu}$ is given by
\begin{equation}
C_{\mu \nu}=-\frac{1}{4} \epsilon^{\alpha \beta \gamma}_{~~~~\mu} \left( R^\sigma_{~\nu \alpha \beta} \nabla_\sigma \nabla_\gamma \psi +2 \nabla_\gamma \psi ~\nabla_\alpha R_{\nu \beta} \right)+(\mu \leftrightarrow \nu).
\end{equation}
There is an identity that $C_{\mu \nu}$ obeys \cite{Jackiw:2003pm},
\begin{equation}
\nabla^\mu C_{\mu \nu}=\frac{1}{8}\psi_{,\nu} {\tilde R}R. \label{identity-J}
\end{equation}
Therefore, by taking the divergence of Eq.~(\ref{NDCD-basic-eq}), we obtain the
following equation,
\begin{equation}
{\tilde R}R=0. \label{CS-cons}
\end{equation}
In other words, any solution of Eq.~(\ref{NDCD-basic-eq}) automatically satisfies the last equation.

In the original paper \cite{Jackiw:2003pm}, $\psi$ is regarded as an external field in a sense that
the action is not variated with respect to $\psi$.
There is no general principle to determine the functional form of $\psi$
and suitable and simple form of $\psi$ is assumed a priori.
However, as is mentioned in Ref.~\cite{Jackiw:2003pm}, even if we take a variation with respect to $\psi$,
we do not get independent equation from Eq.~(\ref{NDCD-basic-eq}).
Indeed, the variation with respect to $\psi$ just yields the constraint (\ref{CS-cons}). 
Just for later convenience, we will treat $\psi$ as a fundamental field.

\subsection{Brief review of the Regge-Wheeler formalism}
Throughout the paper, we consider a static and spherically symmetric spacetime as a background.
Since ${\tilde R}R$ exactly vanishes on this background, 
the background is Schwarzschild spacetime whose metric is given by
\begin{equation}
ds^2=-\left( 1-\frac{r_g}{r} \right) dt^2+{\left( 1-\frac{r_g}{r} \right)}^{-1} dr^2+r^2 \left( d\theta^2+\sin^2 \theta ~d\varphi^2 \right).
\end{equation}
For the CS scalar function, we assume that it takes a form of 
\begin{equation}
\psi =\mu t+\Psi(r),
\end{equation}
where $\Psi(r)$ is an arbitrary function. 
A choice $\Psi=0$ is called canonical embedding and extensively studied in Ref.~\cite{Yunes:2007ss}.

As is well known, the metric perturbations on this background can be decomposed into
odd-type and even-type perturbations according to their transformation properties under
the two-dimensional rotation (the Regge-Wheeler formalism)\cite{Regge:1957td,Zerilli:1970se}.
Furthermore, each perturbation can be decomposed into the sum of spherical harmonics.
Then at the linear order in perturbation equations, or equivalently at the second order
in the perturbative action, the perturbation variables having different $\ell$ and $m$
do not mix each other.
Since detailed explanation of the Regge-Wheeler formalism is provided in Ref.~\cite{Motohashi:2011pw},
we here write down only definition of the notation of the metric perturbations.

For the odd-type perturbations, they can be written as
\begin{eqnarray}
 &  & h_{tt}=0,~~~h_{tr}=0,~~~h_{rr}=0,\\
 &  & h_{ta}=\sum_{\ell, m}h_{0,\ell m}(t,r)E_{ab}\partial^{b}Y_{\ell m}(\theta,\varphi),\\
 &  & h_{ra}=\sum_{\ell, m}h_{1,\ell m}(t,r)E_{ab}\partial^{b}Y_{\ell m}(\theta,\varphi),\\
 &  & h_{ab}=\frac{1}{2}\sum_{\ell, m}h_{2,\ell m}(t,r)\left[E_{a}^{~c}\nabla_{c}\nabla_{b}Y_{\ell m}(\theta,\varphi)+E_{b}^{~c}\nabla_{c}\nabla_{a}Y_{\ell m}(\theta,\varphi)\right]. \label{odd-ab}
\end{eqnarray}
For the even-type perturbations, they can be written as
\begin{eqnarray}
 &  & h_{tt}=A(r)\sum_{\ell, m}H_{0,\ell m}(t,r)Y_{\ell m}(\theta,\varphi),\\
 &  & h_{tr}=\sum_{\ell, m}H_{1,\ell m}(t,r)Y_{\ell m}(\theta,\varphi),\\
 &  & h_{rr}=\frac{1}{B(r)}\sum_{\ell, m}H_{2,\ell m}(t,r)Y_{\ell m}(\theta,\varphi),\\
 &  & h_{ta}=\sum_{\ell, m}\zeta_{\ell m}(t,r)\partial_{a}Y_{\ell m}(\theta,\varphi),\\
 &  & h_{ra}=\sum_{\ell, m}\alpha_{\ell m}(t,r)\partial_{a}Y_{\ell m}(\theta,\varphi),\\ 
 &  & h_{ab}=\sum_{\ell, m} K_{\ell m}(t,r) g_{ab} Y_{\ell m}(\theta,\varphi)+\sum_{\ell, m} G_{\ell m}(t,r) \nabla_a \nabla_b Y_{\ell m}(\theta,\varphi)\,. \label{hab}
\end{eqnarray}
Because of general covariance, not all the metric perturbations are physical 
in the sense that some of them can be set to vanish by using the gauge 
transformation $x^{\mu}\to x^{\mu}+\xi^{\mu}$, where $\xi^{\mu}$ is infinitesimal function.
For the odd-type perturbations, we can completely fix a gauge by imposing a condition $h_2=0$.
This gauge fixing is called Regge-Wheeler gauge.
For the even-type perturbations, complete gauge fixing is achieved by imposing $\zeta=0,~K=0$ and $G=0$. 
We will use these gauge conditions in the calculation of the second order action.

One characteristic feature of CS gravity is that odd-type and even-type perturbations
having the same $(\ell,~m)$ mix through parity violation term, which does not happen for
$f(R)$ gravity theories where parity is preserved.
As a result, we must treat both the odd and the even modes at the same time,
which we will do in the following sections.

\subsection{Second order action for perturbation variables}
In addition to the metric perturbations, we also need to perturb the CS scalar function $\psi$ that appears in the action.
This field must be also decomposed into the spherical harmonics,
\begin{equation}
\delta \psi=\sum_{\ell, m}\delta \psi_{\ell m}(t,r)Y_{\ell m}(\theta,\varphi).
\end{equation}

With these perturbation variables, 
expanding the action (\ref{act1}) to second order (the first order part automatically vanishes because of the background equations) yields the following action,
\begin{eqnarray}
S=\int dt~dr~{\cal L},
\end{eqnarray}
where ${\cal L}$ is written as
\begin{eqnarray}
{\cal L}&=&H_0 (a_1 H_2+a_2 \alpha+a_3 h_0+a_4 H_2'+a_5 \alpha')+b_1 H_1^2+H_1(b_2 h_0+b_3 h_1+b_4 {\dot H_2}+b_5 {\dot \alpha}) \nonumber \\
&&+c_1 h_0^2+c_2 h_0 \alpha+c_3 h_1^2+c_4 h_1 \alpha+c_5 \alpha H_2+c_6 \alpha^2+c_7 \alpha h_0'+c_8\alpha h_1'+c_9 \alpha {\dot h_0}+c_{10} h_1 H_2+c_{11} H_2^2 +c_{12} {\dot \alpha}^2 \nonumber \\
&&+d_1 \delta \psi \left({\dot h_1}-h_0'+\frac{2}{r}h_0 \right) \label{bare-lag}
\end{eqnarray}
Since different $(\ell,~m)$ modes do not mix with each other, we pick up particular $(\ell,~m)$ modes.
Because of spherical symmetry of the background spacetime, the action for $m \neq 0$ modes takes exactly the same form as that for $m=0$,
which enables us to set $m=0$ without loss of generality.
From now on, we abbreviate the subscripts $\ell$ and $m$.
Explicit expressions for the background-dependent coefficients are given in Appendix \ref{ap-a}.
Notice that due to the parity-violating nature of the CS term, 
there appear mixing terms of odd and even perturbations in the action.

This action shows that not all of the variables are dynamical.
Actually, we see that $H_0$ and $H_1$ are auxiliary fields.
Therefore, they can be eliminated from the action by using their equations of motion.
Since $H_0$ appears only linearly, the variation with respect to it gives a constraint among the other fields.
Just for convenience, we introduce a new variable $\beta$ defined by
\begin{equation}
\beta=\alpha+\frac{a_4}{a_5} H_2.
\end{equation}
After this field transformation, the constraint coming from the variation of $H_0$ becomes
algebraic equation for $H_2$. 
Then, we can eliminate $H_2$ by using this constraint.
The variation with respect to $H_1$ gives the equation of motion for $H_1$.
After substituting the constraints and many integration by parts, we end up
with the following Lagrangian density,
\begin{eqnarray}
{\cal L}&=&p_1 h_0' \beta'+p_2 \beta' {\dot h_0}+p_3 {\dot h_0} {\dot \beta}+p_4 h_1' \beta'+p_5 {\dot \beta}^2+p_6 \beta'^2+p_7 h_0 h_1'+p_8 h_0 {\dot \beta}+p_9 h_0 \beta'+p_{10} h_0' \delta \psi+p_{11} h_1 \beta'+p_{12} h_1 {\dot {\delta \psi}} \nonumber \\
&&+p_{13} h_0^2+p_{14} h_0 h_1+p_{15} h_0 \beta+p_{16} h_0 \delta \psi+p_{17} h_1^2+p_{18} h_1 \beta+p_{19}\beta^2. \label{fin-lag}
\end{eqnarray}
Explicit expressions of the background dependent coefficients $p_1,\cdots$ are given in Appendix \ref{ap-p}.

We find that the Lagrangian does not contain time derivative of $h_1$.
This is because we have replaced ${\dot h_1}$ with $h_0'-\frac{2}{r} h_0$ in deriving Eq.~(\ref{fin-lag}). 
This relation holds from the equation of motion for $\delta \psi$, which is derived from Eq.~\eqref{bare-lag}.
Therefore, $h_1$ is an auxiliary field and its equation of motion relates $h_1$ with other fields,
which is given by
\begin{equation}
h_1=\frac{-p_{18}\beta+(p_7'-p_{14})h_0+p_7 h_0'+(p_4'-p_{11})\beta'+p_4 \beta''-p_{12} {\dot {\delta \psi}}}{2p_{17}}.
\end{equation}
Substituting this equation back into Eq.~(\ref{fin-lag}) yields a new Lagrangian which contains only
$h_0,~\delta \psi$ and $\beta$.
The expression of the new Lagrangian is rather long.
Formally, it can be written as 
\begin{equation}
{\cal L}={\cal L}({\dot h_0},{\dot {\delta\psi}},{\dot \beta},h_0',\psi',\beta',\beta'',h_0,\delta\psi,\beta). \label{lag-dy}
\end{equation}
As usual, the kinetic matrix $K_{ij}$ is given by
\begin{equation}
K_{ij}=\frac{\partial^2 {\cal L}}{\partial {\dot q_i} \partial {\dot q_j}},
\end{equation}
where $q_1=h_0,~q_2=\delta\psi$ and $q_3=\beta$.
By the straightforward calculation, we find that the determinant of $K_{ij}$ is given by
\begin{equation}
\det (K_{ij})=-\frac{589824 \pi ^3 \ell^3 (\ell+1)^3 \left(\ell^2+\ell-2\right) M_P^2 (1-A)^2 \Psi'^2}{(2 \ell+1)^3 r^2 A \left(-3 A+\ell^2+\ell+1\right)^2 \left(4
   \left(\ell^2+\ell-2\right) A \Psi'^2+M_P^4 r^2\right)}. \label{det-k}
\end{equation}
This is nonvanishing (A case in which $\Psi'=0$ will be studied later).
Therefore, all $q_i$ are dynamical fields and the Lagrangian (\ref{lag-dy}) cannot be reduced anymore.
Variation with respect to $q_i$ yields the closed set of the evolution equations for these fields.
Since the derived evolution equations involve second time derivatives for
all the fields, we need 6 initial conditions to specify the subsequent time evolution.
Once the solution to those equations is obtained, the evolutions of the other fields
such as $h_1,~H_2,~\alpha,~H_0$ and $H_1$ are uniquely determined by the constraint equations.
This completes the procedures to know the behaviors of all the perturbation variables.
Thus, the problem of solving the perturbation equations of (\ref{NDCD-basic-eq}),
which is highly involved due to mixture of the dynamical and no-dynamical variables,
has been reduced to a much simpler problem. 

The result that we have 6 degrees of freedom for the initial conditions may sound
counterintuitive at a first glance.
For the case of GR, it is well known that we need only four initial conditions.
Since the introduction of the CS term imposes a new constraint (\ref{CS-cons})
which does not exist in GR, one may expect we need less initial conditions
in the nondynamical CS theories.
We think it is the higher derivative terms in the CS term that increase the 
number of initial conditions and the constraint (\ref{CS-cons}) is not strong
enough to reduce the degrees of freedom to less than four.
We shall confirm this point again when we explore dynamical CS theory in the
next section. 

Although there is no apparent mathematical problem for the Lagrangian (\ref{lag-dy}),
physically this Lagrangian is problematic since $\det (K_{ij})$ given by Eq.~(\ref{det-k}) is negative.  
The situation becomes clearer if we construct the Hamiltonian corresponding to this Lagrangian.
The conjugate momenta are defined by
\begin{eqnarray}
&&\pi_0=\frac{\partial {\cal L}}{\partial {\dot h_0}}=\frac{4 p_2 p_{17} \beta'+4p_3 p_{17} {\dot \beta}}{4 p_{17}}, \\
&&\pi_\psi=\frac{\partial {\cal L}}{\partial {\dot {\delta \psi}}}=\frac{p_{12} \left(p_7 h_0'+ \left(p_7'-p_{14} \right)h_0-p_{12} {\dot {\delta \psi}}-p_{18} \beta +(p_4'-p_{11}) \beta'+p_4 \beta'' \right)}{2 p_{17}}, \\
&&\pi_\beta=\frac{\partial {\cal L}}{\partial {\dot \psi}}=p_3 {\dot h_0}+p_8 h_0+2 p_5 {\dot \beta}.
\end{eqnarray}
Since $\det (K_{ij})$ is not zero, the Lagrangian is not singular and we can solve the above equations in terms of the field time derivatives.
The Hamiltonian is then given by
\begin{eqnarray}
H=\int dr~{\cal H}=\int dr~ \left( \pi_0 {\dot h_0}+ \pi_\psi {\dot {\delta \psi}}+\pi_\beta {\dot \beta}-{\cal L}  \right).
\end{eqnarray}
It can be confirmed that the momentum part of the Hamiltonian can be written by
\begin{equation}
{\cal H} =-\frac{p_5}{p_3^2} \pi_0^2+\frac{\pi_0 \pi_\beta}{p_3}-\frac{p_{17}}{p_{12}^2} \pi_\psi^2+\cdots,
\end{equation}
where $\cdots$ represents terms that are not quadratic in $\pi_i$. 
Now it is obvious that the subspace of the Hamiltonian spanned by $\pi_0$ and $\pi_\beta$ yields a negative determinant of the corresponding subkinetic matrix.
This means that the Hamiltonian can take arbitrary negative values by suitably choosing the values of $\pi_0$ and $\pi_\beta$ and hence it is not bounded from below.
To conclude, if $\Psi' \neq 0$, then ghost appears in the action.

Next, let us consider the canonical embedding case where $\Psi'=0$
and start again at Eq.~(\ref{fin-lag}).  
In this case, we can rewrite (\ref{fin-lag}) as
\begin{eqnarray}
{\cal L}&=&p_5 {\dot \beta}^2+p_6 \beta'^2+p_{13}h_0^2+p_{17} h_1^2+p_{19} \beta^2+h_1 \left( p_{18} \beta+(-p_4'+p_{11}) \beta'-p_4 \beta''+p_{12} {\dot {\delta\psi}} \right) \nonumber \\
&&+h_0 \left( (p_{16}-p_{10}')\delta\psi-p_{10}\psi'-p_2 {\dot \beta}'+p_8 {\dot \beta}+p_{15}\beta \right).
\end{eqnarray}
Since no derivatives of $h_0$ nor $h_1$ appear in the new Lagrangian, both variables can
be eliminated by using their equations of motion.
The resulting Lagrangian is given by
\begin{eqnarray}
{\cal L}&=&-\frac{\left( \left(p_{16}-p_{10}'\right) \delta\psi -p_{10} \delta\psi'+p_{15} \beta -p_2 {\dot \beta}'+p_8 {\dot \beta} \right)^2}{4 p_{13}}-\frac{\left(
   \left( p_{11}-p_4' \right) \beta'+p_{12} {\dot {\delta\psi}}+p_{18} \beta -p_4 \beta''\right)^2}{4 p_{17}} \nonumber \\
   &&+p_{19} \beta^2+p_5 {\dot \beta}^2+p_6 \beta'^2.
\end{eqnarray}
The presence of a term ${\dot \beta}'$ prevents us from constructing the corresponding
Hamiltonian straightforwardly.
Because of this, we only consider modes whose wavelength is much shorter than
the length scale of the background curvature radius.
This amounts to an approximation for the fields that we keep only the highest derivative
terms in the Lagrangian.
After the approximation, we have
\begin{equation}
{\cal L}=-\frac{\left( p_{10} \delta\psi' +p_2 {\dot Q} \right)^2}{4 p_{13}}-\frac{\left(p_{12} {\dot {\delta\psi}}-p_4 Q'\right)^2}{4 p_{17}}, 
\end{equation}
where we have defined $Q=\beta'$.
We can then construct the Hamiltonian in a usual way.
The result is given by
\begin{equation}
{\cal H}=\frac{\pi_\psi  \left( p_{12} p_4 Q'-p_{17} \pi_\psi \right)}{p_{12}^2}-\frac{{\pi_Q} \left(p_{10} p_2 \delta\psi'+p_{13} \pi_Q \right)}{p_2^2}.
\end{equation}
Now, let us consider a subspace spanned by $\pi_\psi$ and $\pi_Q$,
with $\psi$ and $Q$ being fixed to zero.
The Hamiltonian reduces to 
\begin{equation}
{\cal H}=-\frac{p_{17}}{p_{12}^2} \pi_\psi^2-\frac{p_{13}}{p_2^2}{\pi_Q}^2.
\end{equation}
Substituting explicit expressions for $p_i$ given in Appendix \ref{ap-p}, 
we find that each background coefficient is given by
\begin{eqnarray}
&&-\frac{p_{17}}{p_{12}^2} =\frac{(2 \ell+1) \left(-3 A+\ell^2+\ell+1\right)^2 \left(4 \left(\ell^2+\ell-2\right) \mu_0^2-M_P^4 r^2 A\right)}{512 \pi \ell \left(\ell^3+2
   \ell^2-\ell-2\right) M_P^2 r^2 \mu_0^2 A^2},\\
&&-\frac{p_{13}}{p_2^2}=\frac{(2 \ell+1) \left( \ell^2+\ell-2\right) M_P^2 r^2 A}{1152 \pi  \ell (\ell+1) (A-1)^2}.   
\end{eqnarray}
The second one is positive definite.
The first one is also positive at the vicinity of the event horizon where $A \approx 0$.
However, it becomes inevitably negative far from the horizon.
In such regions, a ghost appears and the theory becomes problematic.

\section{BH Perturbation in Dynamical Chern-Simons theory}
\label{d-CS}
\subsection{Action and field equations}
Let us now switch on the kinetic term for the scalar field $\psi$ to make $\psi$ a
dynamical field.
For generality, we also add potential term into the action.
Thus, the starting action we study is given by
\begin{equation}
S=\frac{M_P^2}{2} \int d^4x~\sqrt{-g} \left( R-\frac{\xi}{4} \psi {\tilde R}R-\frac{1}{2} {( \nabla \psi )}^2-V(\psi) \right). \label{d-CS-action}
\end{equation}
Notice that $\psi$ in this case is dimensionless.
Correspondingly, a free parameter $\xi$ has the dimension of (length)$^2$.
The gravitational field equations are given by
\begin{equation}
G_{\mu \nu}+\xi C_{\mu \nu}=\frac{1}{2} \partial_\mu \psi \partial_\nu \psi-\frac{1}{2} g_{\mu \nu} \left( \frac{1}{2} {(\partial \psi)}^2+V(\psi) \right). \label{metric-eq-dcs}
\end{equation}
Variation with respect to $\psi$ yields its equation of motion;
\begin{equation}
\nabla_\mu \nabla^\mu \psi-V_\psi-\frac{\xi}{4} {\tilde R}R=0. \label{scalar-eq}
\end{equation}
In the dynamical CS case, taking divergence of Eq.~(\ref{metric-eq-dcs}) 
does not lead to the constraint (\ref{CS-cons}) on the metric variables.
Instead, it just gives Eq.~(\ref{scalar-eq}).
It is clear from the last equation that $\psi$ is dynamical and has
propagating degree of freedom.

Since $\psi$ can source the energy-momentum tensor even on background level, 
the background is not generally Schwarzschild spacetime.
We therefore write the background spacetime as
\begin{equation}
ds^2=-A(r) dt^2+\frac{1}{B(r)} dr^2+r^2 \left( d\theta^2+\sin^2 \theta ~d\varphi^2 \right).
\end{equation}
For the CS scalar function, we write it as 
\begin{equation}
\psi =\Psi(r),
\end{equation}
which is consistent with an assumption that the background solutions are
static and spherically symmetric.
Unlike in the case of nondynamical CS theories, we can not choose
$\Psi$ arbitrarily by hand.
$\Psi$ as well as $A$ and $B$ obey the background equations of motion,
\begin{eqnarray}
&&4 r B'+B \left(r^2 \Psi'^2+4\right)+2 r^2 V-4=0, \label{eq-A} \\
&&4 r B A'+A \left(B \left(4-r^2 \Psi '^2\right)+2 r^2 V-4\right)=0, \label{eq-B}\\
&&r B A' \Psi'+A \left(r B' \Psi'+2 r B \Psi''+4 B \Psi'-2 r V_\psi\right)=0. \label{eq-Psi}
\end{eqnarray}

\subsection{Second order action for perturbation variables}

Putting the metric perturbations and the scalar field perturbation into the action and expanding it to
second order in perturbation, we obtain the following action,
\begin{eqnarray}
S=\int dt~dr~{\cal L},
\end{eqnarray}
where ${\cal L}$ is written as
\begin{eqnarray}
{\cal L}&=&H_0 ({\bar a_1} H_2+{\bar a_2} \alpha+{\bar a_3} \delta \psi+{\bar a_4} H_2'+{\bar a_5} \alpha'+{\bar a_6} \delta \psi'+{\bar a_7} h_0+{\bar a_8} h_0'+{\bar a_9} {\dot h_1}+{\bar a_{10}} h_0''+{\bar a_{11}} {\dot h_1}') \nonumber \\
&&+{\bar b_1} H_1^2+H_1({\bar b_2} {\dot H_2}+{\bar b_3} {\dot \alpha}+{\bar b_4} h_1+{\bar b_5} {\dot h_0}+{\bar b_6} {\dot h_0}'+{\bar b_7} {\ddot h_1}) \nonumber \\
&&+{\bar c_1} H_2^2+H_2 ({\bar c_2} \alpha+{\bar c_3} \delta \psi+{\bar c_4} \delta \psi'+{\bar c_5} h_0+{\bar c_6} h_0'+{\bar c_7} {\dot h_1}) \nonumber \\
&&+{\bar d_1} {\dot \alpha}^2+{\bar d_2} \alpha^2+\alpha ({\bar d_3} h_0+{\bar d_4} h_0'+{\bar d_5} {\dot h_1}+{\bar d_6} {\ddot h_0}+{\bar d_7} {\ddot h_0}'+{\bar d_8} h_{1,ttt}) \nonumber \\
&&+{\bar e_1} \delta \psi^2+{\bar e_2} \delta \psi'^2+{\bar e_3} {\dot {\delta \psi}}^2+\delta \psi ({\bar e_4} h_0+{\bar e_5} h_0'+{\bar e_6} {\dot h_1})+{\bar f_1} h_0^2+{\bar f_2} h_1^2+{\bar f_3} {(h_0'-{\dot h_1})}^2+{\bar f_4} h_0 {\dot h_1}. \label{bare-lag-dcs}
\end{eqnarray}
Explicit expressions of the background dependent coefficients ${\bar a_1},\cdots$ are given in the Aappendix \ref{ap-da}.
As is the case with nondynamical CS gravity, we omitted $\ell$ and $m$
for the multi-pole expansion.

We find that $H_0$ and $H_1$ are again auxiliary fields.
They appear in the action exactly in the same manner as the nondynamical case.
Therefore, in order to reduce the action into the one that contains only dynamical variables,
we can do exactly the same manipulation as the nondynamical case.
Since $H_0$ appears only linearly, the variation with respect to it gives a constraint among the other fields.
Just for convenience, we again introduce a new variable $\beta$ defined by
\begin{equation}
\beta=\alpha+\frac{1}{\bar a_5}( {\bar a_4} H_2+{\bar a_6} \delta\psi+{\bar a_{10}} h_0'+{\bar a_{11}} {\dot h_1}).
\end{equation}
After this field transformation, the constraint coming from the variation of $H_0$ becomes
algebraic equation for $H_2$. 
Then, we can eliminate $H_2$ by using this constraint.
The variation with respect to $H_1$ gives the equation of motion for $H_1$.
After substituting the constraints and many integration by parts, we end up
with the following Lagrangian density,
\begin{eqnarray}
{\cal L}&=&s_1 {( {\ddot h_1}+s_2 {\dot h_0}'+s_3 {\dot h_0})}^2+s_4 {( {\dot \beta}'+s_5 {\dot h_0}'+s_6 {\dot {\delta\psi}}+s_7 {\dot h_0} )}^2+s_8 h_0'^2+s_9 {\dot h_0}^2+s_{10} h_0' {\dot h_1}+s_{11} h_0' \beta'+s_{12} {\dot h_0}{\dot \beta} \nonumber \\
&&+s_{13} h_0' \delta\psi'+s_{14} {\dot h_0} {\dot {\delta\psi}}+s_{15} {\dot h_1}^2+s_{16} \beta' {\dot h_1}+s_{17}\beta'^2+s_{18} {\dot \beta}^2+s_{19} \beta' \delta\psi'+s_{20} {\dot \beta} {\dot {\delta\psi}}+s_{21} \delta\psi'^2+s_{22} {\dot {\delta\psi}}^2+s_{23} h_0 {\dot h_1} \nonumber \\
&&+s_{24} \beta h_0'+s_{25} \delta\psi h_0'+s_{26} \beta {\dot h_1}+s_{27} \delta\psi {\dot h_1}+s_{28} \delta\psi \beta'+s_{29} h_0^2+s_{30} h_0 \beta+s_{31} h_0 \delta\psi+s_{32} h_1^2+s_{33} \beta^2+s_{34} \beta \delta\psi \nonumber \\
&&+s_{35} \delta\psi^2. \label{lag-dcs}
\end{eqnarray}
Since most of the expressions of $s_1,\cdots$ are involved and only
$s_1$ is important for our purpose to show the presence of ghost in the theories, 
we here provide explicit form of $s_1$ only;
\begin{equation}
s_1=-\frac{2\ell (\ell+1) M_P^2 \pi \xi^2 B^{3/2} \Psi'^2}{(2\ell+1) A^{3/2}}.
\end{equation}
This does not vanish if neither $\xi$ nor $\Psi'$ vanishes. 
It is clear from Eq.~(\ref{lag-dcs}) that we need ten initial conditions to specify
the time evolution of perturbation variables.
Taking into account that the dynamical CS scalar has 2 degrees of freedom,
addition of the CS term into the theory therefore brings new 4 degrees of freedom.
This result is consistent with our finding in the previous section that there are 6 
degrees of freedom in the nondynamical CS case because the constraint (\ref{CS-cons}) kills 
2 degrees of freedom.

The presence of the first term which contains the second time derivative squared is problematic 
due to the Ostrogradskii's theorem \cite{Woodard:2006nt}.
To see it in more detail, let us rewrite first term of the Lagrangian (\ref{lag-dcs}) as
\begin{equation}
{\cal L}=s_1 \bigg[ -q^2+2q ( {\ddot h_1}+s_2 {\dot h_0}'+s_3 {\dot h_0}) \bigg]+\cdots. \label{sqq1}
\end{equation}
It can be confirmed that after eliminating $q$ by using its equation of motion, 
this Lagrangian reduces to (\ref{lag-dcs}).
We can implement integration by parts to move one of the second time derivatives acting on $h_1$ 
to $q$, which results in
\begin{equation}
{\cal L}=s_1 \bigg[ -q^2-2{\dot q} {\dot h_1}+2q(s_2 {\dot h_0}'+s_3 {\dot h_0}) \bigg]+\cdots. \label{sqq2}
\end{equation}
Now, the two time derivative term has disappeared and the Lagrangian contains at most first order
in field's time derivative.
It is clear from the last Lagrangian that the subkinetic matrix spanned by ${\dot h_1}$ and ${\dot q}$ has
negative determinant;
\begin{equation}
\begin{array}{|cc|}
\frac{1}{2} \frac{\partial^2 {\cal L}}{\partial {\dot h_1}^2} & \frac{1}{2} \frac{\partial^2 {\cal L}}{\partial {\dot h_1} \partial {\dot q}} \\
\frac{1}{2} \frac{\partial^2 {\cal L}}{\partial {\dot h_1} \partial {\dot q}} & \frac{1}{2} \frac{\partial^2 {\cal L}}{\partial {\dot q}^2} \\
\end{array}
=-s_1^2=-{\left( \frac{2\ell (\ell+1) M_P^2 \pi \xi^2 B^{3/2} \Psi'^2}{(2\ell+1) A^{3/2}} \right)}^2 <0. \label{kine-dcs}
\end{equation}
We therefore conclude that ghost is inevitably present on the BH background in dynamical Chern-Simon gravity. 
On the other hand, if we assume that the CS term in the Lagrangian (\ref{d-CS-action}) is a small correction to GR
and that it should be treated only as perturbations for the modes that are already present in GR,
the number of propagating degrees of freedom is four and the ghost degrees of freedom never appear. 

Eq.~(\ref{kine-dcs}) also shows that presence of the ghost can be avoided if $s_1$ identically vanishes.
This is achieved when either $\xi$ or $\Psi'$ vanishes.
A condition $\xi=0$ leads to a simple system where the canonical scalar field $\psi$ is coupled to GR.
Since studies of the BH perturbation for such a system is already established quite well,
we do not consider this case anymore.
We will consider another case in which $\Psi'=0$ is satisfied in the next subsection.

\subsection{Second order action for special cases with $\Psi'=0$}
This case has been extensively studied in Refs.~\cite{Cardoso:2009pk,Molina:2010fb}
by perturbing the field equations of motion.
It was found that gravitational odd and even modes decouple and effects 
of the CS term only appear in the odd-type metric perturbations.
Here, we analyze the same system by using the second order action and also
confirm the results derived in Refs.~\cite{Cardoso:2009pk,Molina:2010fb}.

Let us start from the background equations for $\Psi'=0$. We set $\Psi=\Psi_0$ and $V(\Psi_0)\equiv V_0$. The Eqs.~\eqref{eq-A}--\eqref{eq-Psi} read
\begin{eqnarray}
&&2 r B'+2B +r^2 V_0-2=0, \label{eqsp-A} \\
&&2 r B A'+A \left(2B +r^2 V_0-2\right)=0, \label{eqsp-B}\\
&&r V_\psi A =0. \label{eqsp-Psi}
\end{eqnarray}
Equation~\eqref{eqsp-Psi} gives $V_\psi=0$, which means that $\Psi$ stays the local extremum of the potential. By substituting Eqs.~\eqref{eqsp-A} to \eqref{eqsp-B}, we obtain $(A/B)'=0$, {\it i.e.}, $A/B={\rm const}>0$. We can absorb this positive constant in the definition of $t$ and normalize $B=A$. As a result, Eq.~\eqref{eqsp-B} yields 
\be A'=-\f{A-1}{r}-\f{rV_0}{2}, \ee
and the solution is given by
\be A(r)=1-\f{r_g}{r}-\f{r^2V_0}{6}, \label{sola} \ee
where we set the integration constant as $r_g$. 
This metric describes BH without rotation nor charge in de Sitter spacetime. 
Assuming a hierarchy $r_g \ll V_0^{-1/2}$,
the solutions for $A=0$ are $r=r_g$ and $r=\sqrt{6/V_0}$, which correspond to the horizon for BH and the horizon for de Sitter expansion, respectively. At these horizons, the speed of the particle reach the speed of light. Thus, we cannot observe $r<r_g$ and $r>\sqrt{6/V_0}$. Below, we consider $r_g<r<\sqrt{6/V_0}$. Note that $0<A<1$ in this region.

Let us proceed to the second order action. For $\Psi'=0$, we can divide Lagrangian as
\be \L = \L_\o + \L_\e, \ee
where $\L_\o$ includes $h_0,~h_1,~\delta \psi$ only and $\L_\e$ includes $H_0,~H_1,H_2,~\alpha$ only.
It has been confirmed that at the action level, 
odd-type metric perturbations and even ones are completely
decoupled and the CS scalar field is solely coupled to the
odd-type metric perturbations.
This decoupling was first found in Refs.~\cite{Cardoso:2009pk,Molina:2010fb} from the analysis of the perturbed field equations.

In the following, we first focus on cases with $\ell \ge 2$ and investigate the odd mode in Sec.~\ref{ssec-odd}
and the even mode in Sec.~\ref{ssec-even}, respectively. 
Since the gauge fixing conditions used so far do not completely fix the gauge for $\ell=0$ and $\ell=1$ cases,
we shall treat them separately.
Sec.~\ref{ssec-mono} is devoted to the $\ell=0$ case, 
where we will find that the perturbations just correspond to the shift of the BH mass. 
We explore perturbations with $\ell=1$ in Sec.~\ref{ssec-di}, which for the stationary
perturbations, provides a solution for slowly rotating BH.

\subsubsection{Odd mode}
\label{ssec-odd}
We can rewrite $\L_\o$ in the following form by completing square and integration by parts, 
\be \L_\o = f_1h_0^2+f_2h_1^2+f_3 \delta \psi^2+\f{r^2f_4}{\ell(\ell+1)A}({\dot {\delta \psi}}^2-A^2\delta \psi'^2)+f_4\mk{h'_0-\dot h_1-\f{2}{r}h_0+f_5\delta \psi}^2. \ee
The explicit forms of background dependent coefficients $f_1,~f_2,\cdots$ are displayed in Appendix \ref{ap-f}.
By using the same trick which we have used in Eqs.~\eqref{sqq1} and \eqref{sqq2}, 
we introduce another field $Q$, and move the spatial derivative from $h_0$ and the time derivative from $h_1$ to $Q$ by integration by parts. We can then solve $h_0$ and $h_1$ and erase them by substituting back the solutions. 
After using the background equation, we reach
\be \L_\o = g_1\psi^2+g_2Q^2+g_3(\dot{\delta\psi}^2-A^2\delta\psi'^2)+g_4(\dot Q^2-A^2Q'^2)+g_5\delta\psi Q \ee
The explicit definitions of $g_1,~g_2,\cdots$ are presented in Appendix \ref{ap-g}. 
In particular, we find that $g_1<0,~g_2<0,~g_3>0,~g_4>0$ and $g_5 \propto \xi$. 
There are neither ghost nor gradient instabilities and two modes propagate with the speed of light. 
Since $g_5 \propto \xi$, we can confirm that the coupling between these two modes disappear in the GR limit $\xi \to 0$. 
The remaining task is to check the mass term of the above Lagrangian density. The mass matrix is
\be M=\left(
\begin{array}{cc}
-g_1 & -g_5/2 \\
-g_5/2 & -g_2
\end{array}
\right).
 \ee
If mass squared takes negative value, the fluctuations are exponentially amplified. 
To avoid such instability of the theory, we need that all eigenvalues of $M$ are negative. 
In other words, we impose ${\rm tr}~M<0$ and $\det M>0$. The first condition is automatically satisfied. 
Second condition yields
\be \ell(\ell+1)\mk{\f{4A-2}{r^2}+V_0}\mk{\f{6(A-1)}{r^2}+V_0}^2\xi^2+[\ell(\ell+1)+r^2V_{\psi\psi}]\mk{\f{\ell(\ell+1)+4(A-1)}{r^2}+V_0}>0 \ee
This inequality constrains the allowed value of $\xi$. 
Just to avoid complexity, we neglect $V_0$ in the following.
We note that the last term of the left hand side is positive for $\ell \ge 2$. 
Therefore, the most strict condition comes from $r=r_g$ and $\ell=2$:
\be |\xi| < \f{r_g^2}{6}\sqrt{1+\f{r_gV_{\psi\psi}}{6}}. \ee
If $\xi$ lies outside this range, one of the mass eigenvalue becomes complex nearby the event horizon.

\subsubsection{Even mode}
\label{ssec-even}
Even mode is governed by the Lagrangian 
\bea
\L_\e &=& v_1\mk{H_1^2+\f{2A}{r^2}\alpha^2+\dot\alpha^2} + v_2H_2^2 - v_3H_2\alpha+H_0\kk{v_4H_2+v_3\alpha+2Av_1\mk{-\f{r}{\ell(\ell+1)}H'_2+\alpha'}} \nonumber \\
&&+2v_1H_1\mk{\f{2r}{\ell(\ell+1)}\dot H_2-\dot\alpha}.
\eea
The explicit expressions of coefficients $v_1,v_2,\cdots$ are provided in Appendix \ref{ap-v}. 
We can derive equation of motion for $H_1$ and solve it for $H_1$. After substituting it, we are left with
\bea
\L_\e &=& \f{2Av_1}{r^2}\alpha^2 + v_2H_2^2 - v_3H_2\alpha + H_0\kk{v_4H_2+v_3\alpha+2Av_1\mk{-\f{r}{\ell(\ell+1)}H'_2+\alpha'}} \nonumber \\
&&+\f{4rv_1}{\ell(\ell+1)}\mk{-\f{r}{\ell(\ell+1)}\dot H_2^2+\dot H_2\dot\alpha}.
\eea
To remove the spatial derivatives from $H_2$ and $\alpha$, we introduce a new field $p$ by
\be p=2Av_1\mk{-\f{r}{\ell(\ell+1)}H_2+\alpha}. \label{def-p}\ee
Taking its spatial derivative, we obtain 
\be 2Av_1\mk{-\f{r}{\ell(\ell+1)}H'_2+\alpha'}=p'+\f{2(rAv_1)'}{\ell(\ell+1)}H_2-2(Av_1)'\alpha. \ee 
We can then erase $H'_2$ and $\alpha'$ from the Lagrangian and obtain the term which has the coupling with $H_0$
\be \L_\e \supset H_0\kk{p'+\mk{v_4+\f{2(rAv_1)'}{\ell(\ell+1)}}H_2+\mk{v_3-2(Av_1)'}\alpha}. \ee
Combined with the constraint equation which is given by variating the above term with respect to $H_0$ and Eq.~(\ref{def-p}), we can solve $H_2$ and $\alpha$ in terms of $p$ and $p'$. Thus, we can erase them and obtain the Lagrangian solely made of $p$
\be \L_\e=w_1\mk{w_2p^2+\dot p^2-A^2p'^2}, \ee
where $w_1$ and $w_2$ are given by
\bea
w_1=\f{2\ell(\ell+1)[\ell(\ell+1)-2]}{M_P^2\pi\ell(\ell+1)A[r^2V_0+6A-2\ell(\ell+1)-2]^2},~~~w_2=\f{2\ell(\ell+1)[\ell(\ell+1)-2]A}{r^2[r^2V_0+6A-2\ell(\ell+1)-2]}.
\eea
In particular, we find that $w_1>0$ and $w_2<0$. 
Therefore, there exists one dynamical mode which propagates with the speed of light 
and there are neither ghost, gradient nor tachyonic instabilities.

\subsubsection{Monopole perturbation: $\ell=0$}
\label{ssec-mono}
By setting $\ell=0$ and using integration by parts, we can write the second order action in the separate form,
\be
\L=\L_\psi+\L_\e,
\ee
where $\L_\psi$ and $\L_\e$ are defined by
\bea
\L_\psi&=&\f{2M_P^2\pi r^2}{A}\mk{\dot{\delta\psi}^2-A^2\delta\psi'^2-AV_{\psi\psi}\delta \psi^2}, \label{mo1} \\
\L_\e&=&M_P^2\pi\kk{(2-r^2V_0)H_2^2+4r(AH_0'H_2+2H_1\dot H_2)}. \label{me1}
\eea
Eq.~\eqref{mo1} suggests that there is one mode which propagates with the speed of light and 
there is no tachyonic instability provided $V_{\psi \psi}>0$. 
On the other hand, taking the variation of Eq.~\eqref{me1} with respect to $H_1$ and $H_2$, we obtain
\bea
&&\dot H_2=0, \label{eommn1} \\
&&(2-r^2V_0)H_2-4H_1+2r(AH_0'-2\dot H_1)=0. \label{eommn2}
\eea
In the monopole perturbation, we have to be careful about the residual gauge degree of freedom 
because the gauge conditions $\alpha=\zeta=G=0$ we have imposed so far hold automatically for $\ell=0$. 
By using the residual gauge, we can set $H_1=0$.
This condition does not fix the gauge completely and still allows gauge mode of a form $H_0=C_0(t)$. 
Imposing this gauge condition, Eq.~\eqref{eommn2} becomes
\be H_2=-\f{2rA}{2-r^2V_0}H_0', \label{h2h0} \ee
and Eq.~\eqref{me1} then gives
\be \L_\e=\f{4M_P^2\pi r^2A^2}{-2+r^2V_0}H_0'^2 \ee
The equation of motion for $H_0$ is therefore
\be \f{\partial}{\partial r}\mk{\f{r^2A^2}{-2+r^2V_0}H_0'}=0. \ee
This differential equation has two solutions.
One of them is $H_0=C_1(t)$, which amounts to the gauge mode and can be set zero. 
We are interested in another solution. Performing integration, we have
\be \f{r^2A^2}{-2+r^2V_0}H_0'=C_2, \ee
where the integration constant $C_2$ is independent of both $t$ and $r$, which can be derived from Eqs.~\eqref{eommn1} and \eqref{h2h0}.
By using the background solution \eqref{sola}, we finally obtain 
\be H_0=\f{2C_2}{rA}.\ee
According to the definition of $H_0$, we notice that $C_2$ corresponds to the mass shift of BH.

\subsubsection{Dipole perturbation: $\ell=1$}
\label{ssec-di}
As is the case for $\ell \ge 2$, we find that Lagrangian takes a separate form $\L=\L_\o+\L_\e$.
For $\L_\e$, we can apply the same procedure for $\ell \ge 2$ as we demonstrated in Sec.~\ref{ssec-even}. As a result, the final Lagrangian becomes total derivative. This fact is consistent with $w_1=0$ for $\ell=1$.

On the other hand, $\L_\o$ for $\ell=1$ is given by
\be \L_\o = \f{2M_P^2\pi r^2}{3A}\kk{\dot{\delta\psi}^2-\delta\psi'^2-A\mk{\f{2}{r^2}+V_{\psi\psi}}\delta \psi^2}+\f{4M_P^2\pi}{3}\kk{\phi^2+2f_5\phi\delta\psi}, \ee
where $f_5$ is given in Appendix \ref{ap-f} and $\phi$ is defined by
\be \phi\equiv h_0'-\dot h_1-\f{2}{r}h_0. \ee
The equation of motion for $h_0$ and $h_1$ are given by
\bea
&&\f{\partial}{\partial r}\mk{\f{\partial \L_\o}{\partial \phi}}+\f{2}{r}\f{\partial \L_\o}{\partial \phi}=0, \label{eomdi1} \\
&&\f{\partial}{\partial t}\mk{\f{\partial \L_\o}{\partial \phi}}=0. \label{eomdi2}
\eea
We can regard above equations as partial differential equations for 
$\partial \L_\o/\partial \phi=\phi+f_5\delta\psi$. 
The solution is given by
\be \phi+f_5\delta\psi=\f{C_3}{r^2}, \label{phipsi} \ee
with an integration constant $C_3$ which does not depend on $t$ nor $r$.
As we mentioned before, the gauge condition $h_2=0$ is automatically satisfied for the $\ell=1$ case.
We then use the residual gauge degree of freedom to set $h_1=0$.
This condition does not fix the gauge completely and still allows gauge mode of a form $h_0=C_4(t) r^2$. 
Hence, Eq.~\eqref{phipsi} implies
\be h_0'-\f{2}{r}h_0+f_5\delta\psi=\f{C_3}{r^2}. \ee
By integrating it once more, we derive the formal solution of $h_0$ in terms of the integral of $\delta\psi$,
\be h_0=-\f{C_3}{3r}-r^2\int dr\f{f_5\delta\psi}{r^2}+C_5 (t) r^2. \label{h0psi} \ee
The last term is a gauge mode and can be set to zero.
Next, we derive the equation of motion for $\delta\psi$,
\be \ddot{\delta\psi}-A^2\delta\psi''-\f{A}{2r}[2(A+1)-r^2V_0]\delta\psi'+\f{A}{r^2}(2+r^2V_{\psi\psi})=\f{2Af_5}{r^2}\mk{\f{C_3}{r^2}-f_5\delta\psi}, \label{dpsi} \ee
where we have used Eq.~\eqref{phipsi} to eliminate $\phi$. 
The last equation is a closed differential equation for $\delta \psi$.
Once the evolution of $\delta \psi$ is obtained by solving this equation,
$h_0$ is determined from Eq.~(\ref{h0psi}).
 
If we consider stationary perturbations for which all the perturbation variables depend only on $r$, 
it provides metric and scalar field configurations for the slowly rotating BH.
Just for simplicity, we hereafter neglect $V_0$ and $V_{\psi\psi}$. 
The resultant equation is 
\be \delta\psi''+\f{2r-r_g}{r(r-r_g)}\delta\psi'-\f{2}{r(r-r_g)}\delta\psi=\f{12C_3r_g\xi}{r^6(r-r_g)}+\f{144r_g^2\xi^2}{r^7(r-r_g)}\delta\psi. \ee
Since we cannot find analytic solutions,
we treat CS coupling $\xi$ as a small parameter and expand $\delta\psi$ in terms of $\xi$,
\be \delta\psi(r)=\sum_{n=0}^{\infty}\xi^{2n+1}\delta\psi^{(2n+1)}(r). \ee
Thus, we can write down the differential equation for each order,
\bea
&&\delta {\psi^{(1)}}''+\f{2r-r_g}{r(r-r_g)}\delta {\psi^{(1)}}'-\f{2}{r(r-r_g)}\delta \psi^{(1)}=\f{12C_3r_g}{r^6(r-r_g)}, \label{dp1} \\ 
&&\delta {\psi^{(3)}}''+\f{2r-r_g}{r(r-r_g)}\delta {\psi^{(3)}}'-\f{2}{r(r-r_g)}\delta \psi^{(3)}=\f{144r_g^2\xi^2}{r^7(r-r_g)}\delta \psi^{(1)},\\
&&\cdots,\\
&&\delta {\psi^{(2n+1)}}''+\f{2r-r_g}{r(r-r_g)}\delta {\psi^{(2n+1)}}'-\f{2}{r(r-r_g)}\delta \psi^{(2n+1)}=\f{144r_g^2\xi^2}{r^7(r-r_g)}\delta \psi^{(2n-1)}.
\eea
Notice that the differential equations are composed only by terms of odd power of $\xi$.
On the other hand, from Eq.~\eqref{h0psi}, $h_0$ is expanded by even power of $\xi$,
\be h_0(r)=\sum_{n=0}^{\infty}\xi^{2n}h_0^{(2n)}(r).\ee
As for $h_0^{(0)}$, it is simply given by the first term of Eq.~(\ref{h0psi}); 
\be h_0^{(0)}=-\f{C_3}{3r}. \ee
This coincides with the Kerr metric expanded to first order in angular momentum $J$ 
provided $C_3=\frac{3J}{4\pi M_P^2}$. 
Therefore, $C_3$ corresponds to the angular momentum of BH.

To obtain the corrections due to the CS term, 
we solve the above differential equation system iteratively. 
This method is originally developed in Refs.~\cite{Yunes:2009hc,Konno:2009kg}. 
First, we solve the differential equation \eqref{dp1} for $\delta\psi^{(1)}$,
\be \delta\psi^{(1)}=D_1(2\tilde r-1)+D_2\kk{2+(2\tilde r-1)\log|1-\tilde r^{-1}|}
-\f{C_3}{12r_g^4\tilde r^4}\kk{120\tilde r^4+10\tilde r^2+10\tilde r+9+60\tilde r^4(2\tilde r-1)\log|1-\tilde r^{-1}|},\ee
where we use dimensionless variable $\tilde r\equiv r/r_g$, and $D_1$ and $D_2$ are the integration constants. 
To avoid the divergence at $\tilde r\to 1$ and at $+\infty$, 
the coefficients for $(2\tilde r-1)$ and $(2\tilde r-1)\log|1-\tilde r^{-1}|$ should vanish. 
This condition fixes $D_1$ and $D_2$ as
\be D_1=0,\quad D_2=\f{5C_3}{r_g^4}. \ee
Hence, we obtain
\be \delta\psi^{(1)}=-\f{C_3}{12r_g^4 \tilde r^4}(10\tilde r^2+10\tilde r+9). \label{p1sol} \ee
By substituting the above solution into the formal solution \eqref{h0psi}, we arrive at the next order of $h_0$,
\be h_0^{(2)}=\f{C_3}{336r_g^5\tilde r^6}(280\tilde r^2+240\tilde r+189). \label{h02sol} \ee
The solutions \eqref{p1sol} and \eqref{h02sol} are consistent with the results in Refs.~\cite{Yunes:2009hc,Konno:2009kg}.

The next order solutions are given by
\bea 
\delta\psi^{(3)}&=&\f{2C_3}{147r_g^8 {\tilde r}^9}(700\tilde r^7+700\tilde r^6+630\tilde r^5+560\tilde r^4+500\tilde r^3+450\tilde r^2+245\tilde r+98), \label{psi-3}\\
h_0^{(4)}&=&\f{2C_3}{63063r_g^9 {\tilde r}^{11}}
\left( 65 (9240 \tilde r^7+7920\tilde r^6+6237\tilde r^5+4928\tilde r^4+3960\tilde r^3+3240\tilde r^2+1617 \tilde r)+38808 \right). \label{h-4}
\eea
The first solution (\ref{psi-3}) was also derived by Ref.~\cite{Yunes:2009hc}.
As far as we know, the second solution (\ref{h-4}) is a new result.
Likewise, we can continue this process iteratively.

\section{Conclusion}
We studied linear perturbations on the static and spherically symmetric background spacetime
both for nondynamical and dynamical Chern-Simons theories.
While the literature carry out the similar analysis under the assumption that the background 
Chern-Simons scalar field vanishes, we did not adopt this assumption.
By eliminating nondynamical variables using their constraint equations, 
we derived the reduced second order action from which a set of closed evolution equations containing 
only dynamical variables are obtained immediately and therefore the number of propagating 
degrees of freedom as well.
We showed that six and ten initial conditions are needed to specify the time evolution of 
perturbations for nondynamical and dynamical Chern-Simons theories, respectively.
We also showed that ghost is present both for the nondynamical case and 
for the dynamical case unless the background Chern-Simons scalar field vanishes,
which therefore puts strong constraint on the general CS theories.
It is also found that if the background scalar field vanishes, ghost degrees of freedom
are killed and all the modes propagate at the speed of light.

\begin{acknowledgments}
This work was supported in part by JSPS Research Fellowships for Young Scientists (HM) and Grant-in-Aid for JSPS Fellows No.~1008477 (TS).
\end{acknowledgments}

\newpage

\appendix

\section{Expressions of $a_1,~a_2,\cdots$.}
\label{ap-a}
\bea
a_1&=&-\frac{2 \pi  \left(\ell^2+\ell+2\right) M_P^2}{2 \ell+1},~~~a_2=\frac{2 \pi  \ell (\ell+1) M_P^2 (A+1)}{2 \ell r+r},~~~a_3=\frac{8 \pi  (\ell-1) \ell (\ell+1) (\ell+2) \Psi'}{(2 \ell+1) r^2} ,\nonumber \\
a_4&=&-\frac{4 \pi  M_P^2 r A}{2 \ell+1},~~~a_5=\frac{4 \pi  \ell (\ell+1) M_P^2 A}{2 \ell+1},~~~b_1=\frac{2 \pi  \ell (\ell+1) M_P^2}{2 \ell+1} ,\nonumber \\
b_2&=&-\frac{8 \pi  (\ell-1) \ell (\ell+1) (\ell+2) \mu}{(2 \ell+1) r^2 A},~~~b_3=-\frac{8 \pi  (\ell-1) \ell (\ell+1) (\ell+2) A \Psi'}{(2 \ell+1) r^2} ,\nonumber \\
b_4&=&\frac{8 \pi  M_P^2 r}{2 \ell+1},~~~b_5=-\frac{4 \pi  \ell (\ell+1) M_P^2}{2 \ell+1},~~~c_1=\frac{2 \pi  (\ell-1) \ell (\ell+1) (\ell+2) M_P^2}{(2 \ell+1) r^2 A} ,\nonumber \\
c_2&=&-\frac{8 \pi  (\ell-1) \ell (\ell+1) (\ell+2) (A-1) \Psi'}{(2 \ell+1) r^3},~~~c_3=-\frac{2 \pi \ell \left( \ell^3+2 \ell^2-\ell-2\right) M_P^2 A}{(2 \ell+1) r^2} ,\nonumber \\
c_4&=&-\frac{8 \pi  \ell \left( \ell^3+2 \ell^2-\ell-2\right) \mu (3 A-1)}{(2 \ell+1) r^3},~~~c_5=-\frac{2 \pi  \ell (\ell+1) M_P^2 (A+1)}{2 \ell r+r} ,\nonumber \\
c_6&=&\frac{4 \pi  \ell (\ell+1) M_P^2 A}{(2 \ell+1) r^2},~~~c_7=-\frac{8 \pi  \ell (\ell+1) \left(\ell^2+\ell-2\right) A \Psi '}{(2 \ell+1) r^2},~~~c_8=\frac{16 \pi  \ell (\ell+1) \left(\ell^2+\ell-2\right) \mu A}{(2 \ell+1) r^2} ,\nonumber \\
c_9&=&-\frac{8 \pi  (\ell-1) \ell (\ell+1) (\ell+2) \mu}{(2 \ell+1) r^2 A},~~~c_{10}=\frac{8 \pi  \ell \left(\ell^3+2 \ell^2-\ell-2\right) \mu}{(2 \ell+1) r^2},~~~c_{11}=\frac{2 \pi  M_P^2}{2 \ell+1} ,\nonumber \\
c_{12}&=&\frac{2 \pi  \ell (\ell+1) M_P^2}{2 \ell+1},~~~d_1=-\frac{48 \pi  \ell (\ell+1) (A-1)}{(2 \ell+1) r^2} .\nonumber \\
\eea

\section{Expressions of $p_1,~p_2,\cdots$.}
\label{ap-p}
\begin{eqnarray}
p_1&=&-\frac{32 \pi  \ell (\ell+1) \left(\ell^2+\ell-2\right) A^2 \Psi '}{(2 \ell+1) r \left(-3 A+\ell^2+\ell+1\right)},~~~p_2=-\frac{32 \pi  (\ell-1) \ell (\ell+1) (\ell+2) \mu_0}{(2 \ell+1) r \left(-3 A+\ell^2+\ell+1\right)}, \nonumber \\
p_3&=&\frac{32 \pi  \ell \left(\ell^3+2 \ell^2-\ell-2\right) \Psi '}{(2 \ell+1) r \left(-3 A+\ell^2+\ell+1\right)},~~~ p_4=\frac{32 \pi  \ell (\ell+1) \left(\ell^2+\ell-2\right) \mu_0 A^2}{(2 \ell+1) r \left(-3 A+\ell^2+\ell+1\right)}, \nonumber \\
p_5&=&\frac{8 \pi  \ell \left(\ell^3+2 \ell^2-\ell-2\right) M_P^2 A}{(2 \ell+1) \left(-3 A+\ell^2+\ell+1\right)^2},~~~p_6=-\frac{8 \pi  \left(\ell^2+\ell-2\right) M_P^2 r B \sqrt{\frac{A}{B}} \left(r B A'+\ell (\ell+1) A\right)}{(2 \ell+1) \left(r B A'+A \left(-2 B+\ell^2+\ell\right)\right)^2}, \nonumber \\
p_7&=&\frac{64 \pi  \ell (\ell+1) \left(\ell^2+\ell-2\right)^2 \mu_0 A \Psi '}{(2 \ell+1) M_P^2 r^3 \left(-3 A+\ell^2+\ell+1\right)},~~~p_8=\frac{16 \pi  \ell \left(\ell^3+2 \ell^2-\ell-2\right) \mu_0 (A+1)}{(2 \ell+1) r^2 A \left(-3 A+\ell^2+\ell+1\right)}, \nonumber \\
p_9&=&\frac{32 \pi  \ell \left(\ell^3+2 \ell^2-\ell-2\right) A \left(-3 A^2+\ell^2+\ell+1\right) \Psi '}{(2 \ell+1) r^2 \left(-3 A+\ell^2+\ell+1\right)^2},~~~p_{10}=\frac{48 \pi  \ell (\ell+1) (A-1)}{(2 \ell+1) r^2}, \nonumber \\
p_{11}&=&-\frac{16 \pi  \ell \left(\ell^3+2 \ell^2-\ell-2\right) \mu_0 A (A+1)}{(2 \ell+1) r^2 \left(-3 A+\ell^2+\ell+1\right)},~~~p_{12}=\frac{48 \pi  \ell (\ell+1) (A-1)}{(2 \ell+1) r^2}, \nonumber \\
p_{13}&=&-\frac{2 \pi  \ell \left(\ell^3+2 \ell^2-\ell-2\right)}{(2 \ell+1) M_P^2 r^4 A^2 \left(-3 A+\ell^2+\ell+1\right)^2} \left(-\left(\ell^2+\ell+1\right) A \left(\left(\ell^2+\ell+1\right) M_P^4 r^2+24 \left(\ell^2+\ell-2\right)
   \mu_0^2\right) \right. \nonumber \\
   &&\left. -96 \left(\ell^2+\ell-2\right) A^4 \Psi ' \left(\Psi '-r \Psi ''\right)+A^3 \left(64 \left(\ell^4+2 \ell^3-\ell-2\right)
   \Psi '^2-32 \left(\ell^4+2 \ell^3-\ell-2\right) r \Psi ' \Psi'' \right. \right. \nonumber \\
   &&\left.\left.-9 M_P^4 r^2\right)+A^2 \left(6 \left(\left(\ell^2+\ell+1\right) M_P^4
   r^2+6 \left(\ell^2+\ell-2\right) \mu_0^2\right)-32 \left(\ell^4+2 \ell^3-\ell-2\right) \Psi'^2\right) \right. \nonumber \\
   &&\left.+4 \left(\ell^2+\ell-2\right) \left(\ell^2+\ell+1\right)^2
   \mu_0^2\right), \nonumber \\
p_{14}&=&\frac{16 \pi  \ell (\ell+1) \left(\ell^2+\ell-2\right)^2 \mu_0 \Psi '}{(2 \ell+1) M_P^2 r^4}, \nonumber \\
p_{15}&=&-\frac{16 \pi  \ell \left(\ell^3+2 \ell^2-\ell-2\right)}{(2 \ell+1) r^3 \left(-3 A+\ell^2+\ell+1\right)^2} \left(2 \left(-3 \left(\ell^2+\ell+1\right) A^2+\left(\ell^2+\ell+1\right)^2 A+3 A^3-\ell^2-\ell-1\right) \Psi ' \right. \nonumber \\
&&\left.-r A (A+1) \left(-3 A+\ell^2+\ell+1\right) \Psi ''\right), \nonumber \\
p_{16}&=&-\frac{96 \pi  \ell (\ell+1) (A-1)}{(2 \ell+1) r^3},~~~p_{17}=-\frac{2 \pi  \ell \left(\ell^3+2 \ell^2-\ell-2\right) A \left(4 \left(\ell^2+\ell-2\right) A \Psi '^2+M_P^4 r^2\right)}{(2 \ell+1) M_P^2 r^4}, \nonumber \\
p_{18}&=&\frac{16 \pi  \ell \left(\ell^3+2 \ell^2-\ell-2\right) \mu_0 \left(-\left(8 \ell^2+8 \ell+11\right) A^2+\left(2 \ell^4+4 \ell^3+7 \ell^2+5 \ell+3\right) A+9
   A^3-\ell^2-\ell-1\right)}{(2 \ell+1) r^3 \left(-3 A+\ell^2+\ell+1\right)^2}, \nonumber \\
p_{19}&=&-\frac{8 \pi  \ell \left(\ell^3+2 \ell^2-\ell-2\right) M_P^2 A \left(-3 \left(\ell^2+\ell+1\right) A^2+\left(\ell^2+\ell+1\right)^2 A+3
   A^3-\ell^2-\ell-1\right)}{(2 \ell+1) r^2 \left(-3 A+\ell^2+\ell+1\right)^3}.
\end{eqnarray}

\section{Expressions of ${\bar a_1},~{\bar a_2},\cdots$.}
\label{ap-da}
\bea
{\bar a_1}&=&-\frac{\pi  M_P^2 \sqrt{A} \left(r^2 (-B) \Psi'^2+2 \left(\ell^2+\ell+2\right)-2 r^2 V\right)}{(2 \ell+1) \sqrt{B}},~~~{\bar a_2}=-\frac{\pi  \ell (\ell+1) M_P^2 \sqrt{A} \left(B \left(r^2 \Psi'^2-4\right)+2 r^2 V-4\right)}{2 \sqrt{B} (2 \ell r+r)} ,\nonumber \\
{\bar a_3}&=&\frac{2 \pi  M_P^2 r^2 \sqrt{A} V_\psi}{(2 \ell+1) \sqrt{B}},~~~{\bar a_4}=-\frac{4 \pi  M_P^2 r \sqrt{A} \sqrt{B}}{2 \ell+1},~~~{\bar a_5}=\frac{4 \pi  \ell (\ell+1) M_P^2 \sqrt{A} \sqrt{B}}{2 \ell+1},~~~{\bar a_6}=\frac{2 \pi  M_P^2 r^2 \sqrt{A} \sqrt{B} \Psi'}{2 \ell+1},\nonumber \\
{\bar a_7}&=&\frac{\pi  \ell (\ell+1) M_P^2 \xi  \left(\Psi' \left(2 \left(2 \left(\ell^2+\ell-3\right)+r^2 V\right)-3 B \left(r^2 \Psi'^2+4\right)\right)+8 r V_\psi\right)}{(2 \ell+1) r^2}, \nonumber \\
{\bar a_8}&=&\frac{\pi  \ell (\ell+1) M_P^2 \xi  \left(\Psi' \left(B \left(3 r^2 \Psi'^2+20\right)-2 r^2 V+4\right)-8 r V_\psi \right)}{2 (2 \ell r+r)}, \nonumber \\
{\bar a_9}&=&\frac{\pi  \ell (\ell+1) M_P^2 \xi  \left(8 r V_\psi-\Psi' \left(B \left(3 r^2 \Psi'^2+4\right)-2 r^2 V+4\right)\right)}{2(2 \ell r+r)},~~~{\bar a_{10}}=-\frac{4 \pi  \ell (\ell+1) M_P^2 \xi  B \Psi'}{2 \ell+1}, \nonumber \\
{\bar a_{11}}&=&\frac{4 \pi  \ell (\ell+1) M_P^2 \xi  B \Psi'}{2 \ell+1},~~~{\bar b_1}=\frac{2 \pi  M_P^2 \sqrt{B} \left(-r^2 B \Psi'^2+\ell^2+\ell\right)}{(2 \ell+1) \sqrt{A}},~~~{\bar b_2}=\frac{8 \pi  M_P^2 r \sqrt{B}}{(2 \ell+1) \sqrt{A}},\nonumber \\
{\bar b_3}&=&-\frac{4 \pi  \ell (\ell+1) M_P^2 \sqrt{B}}{(2 \ell+1) \sqrt{A}},~~~{\bar b_4}=-\frac{4 \pi  \ell (\ell+1) \left(\ell^2+\ell-2\right) M_P^2 \xi  B \Psi'}{(2 \ell+1) r^2},~~~{\bar b_5}=-\frac{8 \pi  \ell (\ell+1) M_P^2 \xi  B \Psi'}{(2 \ell+1) r A},\nonumber \\
{\bar b_6}&=&\frac{4 \pi  \ell (\ell+1) M_P^2 \xi  B \Psi'}{(2 \ell+1) A},~~~{\bar b_7}=-\frac{4 \pi  \ell (\ell+1) M_P^2 \xi  B \Psi'}{(2 \ell+1) A},~~~{\bar c_1}=\frac{\pi  M_P^2 \sqrt{A} \left(r^2 B \Psi'^2-r^2 V+2\right)}{(2 \ell+1) \sqrt{B}},\nonumber \\
{\bar c_2}&=&-\frac{\pi  \ell (\ell+1) M_P^2 \sqrt{A} \left(B \left(r^2 \Psi'^2+4\right)-2 r^2 V+4\right)}{2 \sqrt{B} (2 \ell r+r)},~~~{\bar c_3}=-\frac{2 \pi  M_P^2 r^2 \sqrt{A} V_\psi}{(2 \ell+1) \sqrt{B}},\nonumber \\
{\bar c_4}&=&-\frac{2 \pi  M_P^2 r^2 \sqrt{A} \sqrt{B} \Psi'}{2 \ell+1},~~~{\bar c_5}=-\frac{\pi  \ell (\ell+1) M_P^2 \xi  \Psi' \left(B \left(r^2 \Psi'^2-12\right)-2 r^2 V+4\right)}{(2 \ell+1) r^2},\nonumber \\
{\bar c_6}&=&\frac{\pi  \ell (\ell+1) M_P^2 \xi  \Psi' \left(B \left(r^2 \Psi'^2-12\right)-2 r^2 V+4\right)}{2 (2 \ell r+r)},~~~{\bar c_7}=\frac{\pi  \ell (\ell+1) M_P^2 \xi  \Psi' \left(B \left(12-r^2 \Psi'^2\right)+2 r^2 V-4\right)}{2 (2 \ell r+r)},\nonumber \\
{\bar d_1}&=&\frac{2 \pi  \ell (\ell+1) M_P^2 \sqrt{B}}{(2 \ell+1) \sqrt{A}},~~~{\bar d_2}=\frac{2 \pi  \ell (\ell+1) M_P^2 \sqrt{A} \sqrt{B} \left(r^2 B \Psi'^2+2\right)}{(2 \ell+1) r^2},\nonumber \\
{\bar d_3}&=&\frac{\pi  \ell (\ell+1) M_P^2 \xi  \Psi' \left(B \left(\left(\ell^2+\ell-2\right) r^2 \Psi'^2-4 \left(3 \ell^2+3 \ell-2\right)\right)-2
   \left(\ell^2+\ell-2\right) \left(r^2 V-2\right)\right)}{(2 \ell+1) r^3},\nonumber \\
{\bar d_4}&=&\frac{8 \pi  \ell (\ell+1) M_P^2 \xi  B \Psi'}{(2 \ell+1) r^2},~~~{\bar d_5}=-\frac{4 \pi  \ell^2 (\ell+1)^2 M_P^2 \xi  B \Psi'}{(2 \ell+1) r^2},~~~{\bar d_6}=-\frac{8 \pi  \ell (\ell+1) M_P^2 \xi  B \Psi'}{(2 \ell+1) r A}, \nonumber \\
{\bar d_7}&=&\frac{4 \pi  \ell (\ell+1) M_P^2 \xi  B \Psi'}{(2 \ell+1) A},~~~{\bar d_8}=-\frac{4 \pi  \ell (\ell+1) M_P^2 \xi  B \Psi'}{(2 \ell+1) A},~~~{\bar e_1}=-\frac{2 \pi  M_P^2 \sqrt{A} \left(\ell^2+\ell+r^2 V_{\psi \psi} \right)}{(2 \ell+1) \sqrt{B}}, \nonumber \\
{\bar e_2}&=&-\frac{2 \pi  M_P^2 r^2 \sqrt{A} \sqrt{B}}{2 \ell+1},~~~{\bar e_3}=\frac{2 \pi  M_P^2 r^2}{(2 \ell+1) \sqrt{A} \sqrt{B}},~~~{\bar e_4}=\frac{4 \pi  \ell (\ell+1) M_P^2 \xi  \left(B \left(r^2 \Psi'^2-12\right)-2 r^2 V+12\right)}{(2 \ell+1) r^3}, \nonumber \\
{\bar e_5}&=&-\frac{2 \pi  \ell (\ell+1) M_P^2 \xi  \left(B \left(r^2 \Psi'^2-12\right)-2 r^2 V+12\right)}{(2 \ell+1) r^2},~~~{\bar e_6}=\frac{2 \pi  \ell (\ell+1) M_P^2 \xi  \left(B \left(r^2 \Psi'^2-12\right)-2 r^2 V+12\right)}{(2 \ell+1) r^2}, \nonumber \\
{\bar f_1}&=&\frac{\pi  \ell (\ell+1) M_P^2 \left(B \left(4-r^2 \Psi'^2\right)+2 \left(\ell^2+\ell-2\right)\right)}{(2 \ell+1) r^2 \sqrt{A} \sqrt{B}},~~~{\bar f_2}=-\frac{2 \pi  \ell \left(\ell^3+2 \ell^2-\ell-2\right) M_P^2 \sqrt{A} \sqrt{B}}{(2 \ell+1) r^2}, \nonumber \\
{\bar f_3}&=&\frac{2 \pi  \ell (\ell+1) M_P^2 \sqrt{B}}{(2 \ell+1) \sqrt{A}},~~~{\bar f_4}=\frac{8 \pi  \ell (\ell+1) M_P^2 \sqrt{B}}{\sqrt{A} (2 \ell r+r)}
\eea

\section{Expressions of $f_1,~f_2,\cdots$.}
\label{ap-f}
\bea
f_1&=&\f{2\ell(\ell+1)[\ell(\ell+1)-2]M_P^2\pi}{(2\ell+1)r^2A} ,\nonumber \\
f_2&=&\f{2(\ell-1)\ell(\ell+1)(\ell+2)M_P^2\pi A}{(2\ell+1)r^2} ,\nonumber \\
f_3&=&-\f{2M_P^2\pi[\ell(\ell+1)(1+f_5^2)+r^2V_{\psi\psi}]}{2\ell+1} ,\nonumber \\
f_4&=&\f{2M_P^2\pi \ell(\ell+1)}{2\ell+1} ,\nonumber \\
f_5&=&\xi\mk{\f{6(A-1)}{r^2}+V_0}.
\eea

\section{Expressions of $g_1,~g_2,\cdots$.}
\label{ap-g}
\bea
g_1&=&f_3 ,\nonumber \\
g_2&=&-\f{2M_P^2\pi\ell(\ell+1)[\ell(\ell+1)+4(A-1)+r^2V_0]}{(2\ell+1)[\ell(\ell+1)-2]} ,\nonumber \\
g_3&=&\f{2M_P^2\pi r^2}{(2\ell+1)A} ,\nonumber \\
g_4&=&\f{2M_P^2\pi r^2\ell(\ell+1)}{(2\ell+1)[\ell(\ell+1)-2]A} ,\nonumber \\
g_5&=&\f{4M_P^2\pi\ell(\ell+1)f_5}{2\ell+1}.
\eea

\section{Expressions of $v_1,~v_2,\cdots$.}
\label{ap-v}
\bea
v_1&=&\f{2M_P^2\pi \ell(\ell+1)}{2\ell+1} ,\nonumber \\
v_2&=&-\f{M_P^2\pi(r^2V_0-2)}{2\ell+1} ,\nonumber \\
v_3&=&-\f{M_P^2\pi \ell(\ell+1)[r^2V_0-2(A+1)]}{(2\ell+1)r} ,\nonumber \\
v_4&=&\f{2M_P^2\pi [r^2V_0-\ell(\ell+1)-2]}{2\ell+1}.
\eea

\end{document}